\documentclass[11pt]{article}
\usepackage{amsmath,amssymb}
\makeatletter
\@addtoreset{equation}{section}
\makeatother

\def\crampest{\medmuskip = 1mu plus 1mu minus 1mu}

\def\ben{\begin{equation}}
\def\een{\end{equation}}

  \let\n=\nu  \let\p=\pi

\let\C=\Chi

\def\nn{\nonumber} \def\bd{\begin{document}} \def\ed{\end{document}}
\def\ds{\documentstyle} \let\fr=\frac \let\bl=\bigl \let\br=\bigr
\let\Br=\Bigr \let\Bl=\Bigl
\let\bm=\bibitem
\let\na=\nabla
\let\pa=\partial \let\ov=\overline
\newcommand{\be}{\begin{equation}}
\newcommand{\ee}{\end{equation}}
\def\ba{\begin{array}}
\def\ea{\end{array}}
\def\ft#1#2{{\textstyle{\frac{\scriptstyle #1}{\scriptstyle #2} } }}
\def\fft#1#2{{\frac{#1}{#2}}}
\def\del{\partial}
\def\vp{\varphi}
\def\sst#1{{\scriptscriptstyle #1}}
\def\oneone{\rlap 1\mkern4mu{\rm l}}
\def\td{\tilde}
\def\wtd{\widetilde}
\def\ie{{\it i.e.\ }}
\def\dalemb#1#2{{\vbox{\hrule height .#2pt
        \hbox{\vrule width.#2pt height#1pt \kern#1pt
                \vrule width.#2pt}
        \hrule height.#2pt}}}
\def\square{\mathord{\dalemb{6.8}{7}\hbox{\hskip1pt}}}
\newcommand{\ho}[1]{$\, ^{#1}$}
\newcommand{\hoch}[1]{$\, ^{#1}$}
\newcommand{\bea}{\begin{eqnarray}}
\newcommand{\eea}{\end{eqnarray}}
\newcommand{\ra}{\rightarrow}
\newcommand{\lra}{\longrightarrow}
\newcommand{\Lra}{\Leftrightarrow}
\newcommand{\bp}{\tilde \beta^\prime}
\newcommand{\tr}{{\rm tr} }
\newcommand{\Tr}{{\rm Tr} }
\def\0{{\sst{(0)}}}
\def\1{{\sst{(1)}}}
\def\2{{\sst{(2)}}}
\def\3{{\sst{(3)}}}
\def\4{{\sst{(4)}}}
\def\5{{\sst{(5)}}}
\def\6{{\sst{(6)}}}
\def\7{{\sst{(7)}}}
\def\8{{\sst{(8)}}}
\def\n{{\sst{(n)}}}
\def\cA{{{\cal A}}}
\def\cB{{{\cal B}}}
\def\cF{{{\cal F}}}
\def\cH{{{\cal H}}}
\def\tV{\widetilde V}
\def\tW{\widetilde W}
\def\tH{\widetilde H}
\def\tE{\widetilde E}
\def\tF{\widetilde F}
\def\tA{\widetilde A}
\def\im{{{\rm i}}}
\def\tY{{{\wtd Y}}}
\def\ep{{\epsilon}}
\def\vep{{\varepsilon}}
\def\bD{{{\bar D}}}
\def\R{{{\mathbb R}}}
\def\C{{{\mathbb C}}}
\def\H{{{\mathbb H}}}
\def\CP{{{\mathbb C}{\mathbb P}}}
\def\RP{{{\mathbb R}{\mathbb P}}}
\def\Z{{{\mathbb Z}}}
\def\bA{{{\mathbb A}}}
\def\bB{{{\mathbb B}}}
\def\bC{{{\mathbb C}}}
\def\bD{{{\mathbb D}}}
\def\bE{{{\mathbb E}}}
\def\bZ{{{\mathbb Z}}}
\def\Re{{{\frak{Re}}}}
\def\Im{{{\frak{Im}}}}
\def\cosec{{\,\hbox{cosec}\,}}
\def\Gm{{\Gamma_{\!\! -}}}
\def\Gp{{\Gamma_{\!\! +}}}
\def\stan{{standard }}
\def\nonstan{{supernumerary }}
\def\p{{\partial}}
\def\kdel#1{{\fft{\del}{\del#1}}}

\def\bog{{Bogomolny }}
\def\om{{\omega}}

\newcommand{\tamphys}{\it George and Cynthia Woods Mitchell  Institute
for Fundamental Physics and Astronomy,\\
Texas A\&M University, College Station, TX 77843, USA}

\newcommand{\auth}{
G.W. Gibbons\hoch{\dagger}, H. L\"u\hoch{\ddagger,\star} and
C.N. Pope\hoch{\ddagger,\dagger}
}

\begin{document}

\begin{flushright}
\hfill{
DAMTP-2009-20\ \ \ \ \ \ MIFP-09-09}\\
%\hfill{
%\bf hep-th/yymmnnn}
\end{flushright}

\begin{center}

{\large {\bf Einstein Metrics on Group Manifolds and Cosets
             
}}

\vspace{25pt}

\auth

\vspace{10pt}

\hoch{\dagger}{DAMTP, Centre for Mathematical Sciences,
 Cambridge University,\\  Wilberforce Road, Cambridge CB3 OWA, UK}

\vspace{10pt}
\hoch{\ddagger}{\tamphys}

\vspace{10pt}
\hoch{\star}{\it Division of Applied Mathematics and Theoretical
Physics,\\
China Institute for Advanced Study,\\
Central University of Finance and Economics, Beijing, 100081, China
}

\vspace{25pt}

\underline{ABSTRACT}

\end{center}

   It is well known that every compact simple 
group manifold $G$ admits a bi-invariant 
Einstein metric, invariant under $G_L\times G_R$.  Less well known is that
every compact simple group manifold except $SO(3)$ and $SU(2)$ admits 
at least one more homogeneous
Einstein metric, invariant still under $G_L$ but with some, or all, of the
right-acting symmetry broken.  ($SO(3)$ and $SU(2)$ are 
exceptional in admitting only
the one, bi-invariant, Einstein metric.)  In this paper, we look for
Einstein metrics on three relatively low dimensional examples, namely
$G=SU(3)$, $SO(5)$ and $G_2$.  For $G=SU(3)$, we find just the two
already known inequivalent Einstein metrics.  For $G=SO(5)$, we find
four inequivalent Einstein metrics, thus extending previous results where
only two were known.  For $G=G_2$ we find six inequivalent Einstein metrics,
which extends the list beyond the previously-known two examples.  We also
study some cosets $G/H$ for the above groups $G$.  In particular, for 
$SO(5)/U(1)$ we find, depending on the embedding of the $U(1)$,
generically two, with exceptionally one or three, Einstein metrics.   
We also find a pseudo-Riemannian Einstein metric of signature $(2,6)$ on
$SU(3)$, an Einstein metric of signature $(5,6)$ on $G_2/SU(2)_{\rm diag}$,
and an Einstein metric of signature $(4,6)$ on $G_2/U(2)$.
Interestingly, there are no Lorentzian Einstein metrics among our examples.

\vspace{15pt}

\thispagestyle{empty}

\pagebreak
\setcounter{page}{1}

\tableofcontents

\addtocontents{toc}{\protect\setcounter{tocdepth}{2}}

%%%%%%%%%%%%%%%%%%%%%%%%%%%%%%%%%%%%%%%%

\section{Introduction}

    Finding Einstein metrics on compact spaces is a subject of 
considerable mathematical interest.  It is also of importance in physics,
most notably in the compactification of the extra dimensions in supergravity,
string theory or M-theory backgrounds.  A class of Einstein spaces that
was much studied in the 1980's, prior to the rise of string theory and
the consequent emphasis on Ricci-flat Calabi-Yau compactifications, comprised 
compact homogeneous spaces $G/H$, admitting a transitive group action under
$G$.  Even within the framework of string theory or M-theory, such 
compactifications still have an important r\^ole to play, for example in the
AdS/CFT correspondence.

  Motivated by this, we have looked in a somewhat broader context at the
question of the existence of Einstein metrics on compact homogeneous
spaces $G/H$, both for the case where $H$ is some proper subgroup of $G$ and
also for the case that $H$ is the identity, in which case the space is just
the group manifold $G$ itself.  We investigate the first few low-dimensional
examples of compact simple groups $G$.  The first non-trivial example, for
which the group manifold admits a second Einstein metric, is $SU(3)$.  We
focus on $SU(3)$, $SO(5)$ and $G_2$ in our discussions.   

   It is well known that if $g$ denotes a group element in $G$ then the
bi-invariant metric $\tr(g^{-1}\, dg)^2$ is necessarily Einstein, since
all symmetric 2-index tensors that are invariant under $G_L\times G_R$ must
be constant multiples of one another.  However, this does not exclude the
possibility that there could exist further, inequivalent, 
homogeneous Einstein metrics, invariant still under the action of $G_L$, but
with some or all of the $G_R$ symmetry broken.  In fact, it has been shown
by D'Atri and Ziller \cite{datzil} 
that every compact simple group except $SO(3)$ and $SU(2)$
admits at least one such additional Einstein metric.
Such metrics can be 
constructed as follows.  First, we define the left-invariant 1-forms
$\sigma_a$ on $G$:
%%%%%
\be
g^{-1}\, dg = \sigma_a\, T^a\,,
\ee
%%%%%
where $T^a$ are the generators of the Lie algebra of $G$.  Then, the most
general left-invariant metric on $G$ can be written as
%%%%%
\be
ds^2= x_{ab}\, \sigma_a\, \sigma_b\,,\label{sqmet}
\ee
%%%%%
where $x_{ab}$ is a constant symmetric 
``squashing matrix.''  The D'Atri-Ziller examples are obtained by 
rescaling the bi-invariant metric, 
for which by a suitable choice for the basis $\sigma_a$ 
one may take $x_{ab}=\delta_{ab}$,
along a suitably chosen subgroup.

   In principle the problem of looking for Einstein metrics within the
class (\ref{sqmet}) is a purely mechanical one; first one computes the 
Ricci tensor $R_{ab}$
as a function of $x_{ab}$, and then one solves the algebraic
equations resulting from imposing the Einstein condition $R_{ab}=\lambda\,
g_{ab}$.\footnote{If the metric is to be Riemannian, then $x_{ab}$ 
should be positive definite.  It can sometimes happen that pseudo-Riemannian
solutions arise in which $x_{ab}$ is non-singular but indefinite.}
  In practice, however, the technical difficulties of solving the
equations for the $\ft12 d(d+1)$ independent components of $x_{ab}$,
where $n=\hbox{dim}(G)$, can become insurmountable.  One option, which is
the one we shall follow in this paper, is to make some simplifying assumptions
about the structure of $x_{ab}$, in which many of the components are set
to zero, and, possibly, sets of symmetry-related non-zero components are
set equal.  Thus the general idea is to try various restricted ans\"atze for
the coefficients $x_{ab}$, motivated by the symmetries of the situation.

   By following such a strategy, we succeed in finding four inequivalent 
Riemannian Einstein 
metrics on the 10-dimensional group manifold $SO(5)$, of which two appear 
to be new.   We find six inequivalent Riemannian Einstein metrics
on the 14-dimensional
group manifold $G_2$, of which four appear to be new.  
We also find a pseudo-Riemannian Einstein metric of 
signature $(2,6)$ on $SU(3)$, in addition to the two known Riemannian 
Einstein metrics.

  An important question that arises when a candidate ``new'' Einstein 
metric is found on a given space is whether it is genuinely inequivalent
to previously-obtained metrics.  This may not necessarily be easy to
see directly, since it might be that some non-trivial change to the 
basis $\sigma_a$ would
be required in order to reveal the equivalence of two metrics.  Although
it might, therefore, be quite tricky to demonstrate that two ostensibly
different metrics are actually equivalent, the inverse question can often
be easily settled.  We may consider invariant (\ie dimensionless) 
quantities, built from the scalar curvature invariants and the 
magnitude of the volume form of the
manifold.  If such an invariant takes different values for two metrics, then
those metrics are definitely inequivalent.  Two such invariants that
we find useful, in this regard, are
%%%%%
\be
I_1= \lambda^{d/2}\, V\,,\qquad I_2= |\hbox{Riem}|^2\, \lambda^{-2}\,,
\ee
%%%%%
where $\lambda$ is the Einstein constant ($R_{ab}=\lambda\, g_{ab}$), 
$|\hbox{Riem}|^2= R_{abcd}\, R^{abcd}$, $V$ is the magnitude of the 
volume form of the manifold,
and $d$ is its dimension.  By comparing the values of either or both
of these invariants for ostensibly different Einstein metrics on a given 
manifold, one may quickly and unambiguously establish inequivalence, in the
event of unequal values, 
whilst if the invariants take the same values for two metrics this allows
one to focus on the these cases for closer examination.

   It is also of interest to look for Einstein metrics on the homogeneous
spaces $G/H$.  This problem has been studied extensively in the mathematics
literature, and also, in dimensions such as 7 that are particularly 
relevant for Kaluza-Klein compactifications, in the physics literature too.
We discuss some examples where $G$ is $SU(3)$, $SO(5)$ or $G_2$. 
In particular, we find new Einstein metrics on $SO(5)/U(1)$, and we also find a 
pseudo-Riemannian Einstein metric of signature $(6,5)$ on 
$G_2/SU(2)_{\rm diag}$, and one of signature $(4,6)$ on
$G_2/U(2)$.  Surprisingly, perhaps, we find no Einstein
metrics of Lorentzian signature $(1,n)$ on any of the group manifolds or
cosets considered in this paper.  We shall comment further on this in the
conclusion.

\section{$SU(3)$ and $SU(3)/SO(3)_{\rm maximal}$}

  We identify the Lie algebra of $SU(3)$, denoted by $\mathfrak{su}(3)$, with
traceless Hermitean $3\times 3$ matrices $T^A{}_B$, and hence
the left-invariant 1-forms $L_A{}^B$ are complex valued, with
$L_A{}^A=0$ and $(L_A{}^B)^\dagger = L_B{}^A$.  They satisfy the algebra
%%%%%
\be
dL_A{}^B = \im\, L_A{}^C\wedge L_C{}^B\,.
\ee
%%%%%

   It is convenient to decompose the $\mathfrak{su}(3)$ algebra with respect to
its maximal $\mathfrak{so}(3)$ subalgebra:
%%%%%
\bea
K_1 &=& L_2{}^3 + L_3{}^2\,,\qquad
K_2= L_3{}^1 + L_1{}^3\,,\qquad
K_3= L_1{}^2 + L_2{}^1\,,\nn\\
K_4 &=& L_1{}^1 - L_2{}^2\,,\qquad
K_5= \fft1{\sqrt3}\, (L_1{}^1 + L_2{}^2 - 2 L_3{}^3)\,,\nn\\
H_1 &=& \im\, (L_2{}^3 - L_3{}^2)\,,\qquad
H_2= \im\, (L_3{}^1 - L_1{}^3)\,,\qquad
H_3= \im\, (L_1{}^2 - L_2{}^1)\,.
\eea
%%%%%
The subalgebra $\mathfrak{so}(3)_{\rm maximal}$ 
is generated by $H_1$, $H_2$ and $H_3$, and the $K_i$ transform as a
$\bf 5$ under $\mathfrak{so}(3)_{\rm maximal}$.

\subsection{Einstein metrics on $SU(3)$}

   We consider metrics of the form
%%%%%
\be
ds_8^2 = x_1\, (K_1^2 + K_2^2 + K_3^2) + x_2\, K_4^2 + x_3\, K_5^2 +
          x_4\, (H_1^2 + H_2^2 + H_3^2)\,.\label{su3mets}
\ee
%%%%%
We find two inequivalent Riemannian Einstein metrics, given by
%%%%%
\bea
(x_1,x_2,x_3,x_4) &=& (1,1,1,1)\,, \qquad \lambda = \fft{3}{4}\,,
\qquad \qquad |\hbox{Riem}|^2/\lambda^2 = 8\,,\nn\\
(x_1,x_2,x_3,x_4) &=& (11,11,11,1)\,,\qquad \lambda = \fft{63}{484}\,,
\qquad\qquad |\hbox{Riem}|^2/\lambda^2 = \fft{764}{63}\,.
\eea
%%%%%
The first is the bi-invariant metric.

   We also find a third Einstein metric with the class (\ref{su3mets}),
which has indefinite signature $(2,6)$.  This is given (up to scaling) by
%%%%%
\be
(x_1,x_2,x_3,x_4)= \Big(x_1, -\fft{(1-x_1)(1-5 x_1)}{5 x_1},
 -\fft{(1-x_1)(1-5 x_1)}{5 x_1}, 1\Big)\,,\label{x1234}
\ee
%%%%%
where $x_1$ is the real root of the cubic equation
%%%%%
\be
85 x_1^3 - 29 x_1^2 + 27 x_1 - 3=0\,.
\ee
%%%%%
Since this root is given approximately by $x_1\approx 0.12130$, it follows
from (\ref{x1234}) that there will be two timelike directions in the metric
(\ref{su3mets}) in this case.  The Einstein constant is positive, given by
%%%%%
\be
\lambda = \fft{3(1-x_1)(10 x_1-1)}{20 x_1^2\, (1-5 x_1)} \approx 4.848\,.
\ee
%%%%%

\subsection{The five-dimensional coset $SU(3)/SO(3)_{\rm maximal}$}

  $SU(3)$ acts on $SU(3)/SO(3)_{\rm maximal}$, with $SO(3)_{\rm maximal}$
as a stabiliser.  Since the $K_i$ span the tangent space of the coset,
any $SU(3)$-invariant metric must necessarily be invariant under the
 $SO(3)_{\rm maximal}$ subgroup.  As noted above, the $K_i$ transform
as a $\bf 5$ under  $SO(3)_{\rm maximal}$.  This has a unique 
(up to overall scaling) quadratic invariant, and hence the unique
$SU(3)$-invariant metric on the coset  $SU(3)/SO(3)_{\rm maximal}$ is
given by
%%%%%
\be
ds_5^2 = K_1^2 + K_2^2 + K_3^2 + K_4^2 + K_5^2\,.
\ee
%%%%%
Since  $SU(3)/SO(3)_{\rm maximal}$ is a symmetric space, this
 metric is Einstein, and it is easy to see that
%%%%%
\be
\lambda = \fft{3}{2}\,.
\ee
%%%%%
The metric has no Killing spinors, \ie solutions of $\nabla_a\eta =
\im/2\, \sqrt{\lambda/(d-1)}\, \Gamma_a\eta$ where $\Gamma_a$ are the
Dirac matrices, obeying the Clifford algebra $\{\Gamma_a,\Gamma_b\}=2g_{ab}$, 
and in fact it does not admit a spin structure (see, for example, 
\cite{hep-th/0210297}).

   A convenient coordinatisation of the symmetric space 
$SU(3)/SO(3)$ can be given by defining the coset representative
%%%%%
\be
{\cal V} ={\cal V}_1\, {\cal V}_2\,,\qquad
{\cal V}_1 = e^{\im x \lambda_1} e^{\im y \lambda_4} e^{\im z \lambda_6}\,,
\qquad
{\cal V}_2= e^{\im \td\phi_1\, \lambda_3}\, e^{\im\td\phi_2\,\sqrt3\, 
\lambda_8}\,,
\ee
%%%%%
where $\lambda_i$ are the standard Gell-Mann generators for 
$\mathfrak{su}(3)$,  and 
%%%%%%
\be
d{\cal V}\, {\cal V}^{-1} = \im (P_1\, \lambda_1 + P_2\, \lambda_4 + P_3\, 
  \lambda_6 + P_4\, \lambda_3 + P_5\, \lambda_8 +
   Q_1\, \lambda_2 + Q_2\, \lambda_5 + Q_3\, \lambda_7 )\,.
\ee
%%%%%
The metric on $SU(3)/SO(3)_{\rm maximal}$ is then given by
%%%%%
\be
ds_5^2 = P_1^2 + P_2^2 + P_3^2 +P_4^2 +P_5^2\,.
\ee
%%%%%
Defining new azimuthal coordinates
%%%%%
\be
\phi_1= \td\phi_1 + \td\phi_2\,,\qquad 
\phi_2= \td\phi_1 - 3\td\phi_2\,,
\ee
%%%%%
we find that the basis of 1-forms for the coset is given by
%%%%%
\bea
P_1 &=& dx + \ft12 \sin y\, \sin 2z\,\, d\phi_2\,,\nn\\
P_2 &=& \cos x\, dy - \ft12 \sin x\, \cos y\, \sin 2z\,\, d\phi_2\,,\nn\\
P_3 &=&  \cos y\, \Big( \cos x\, dz + \ft12
            \sin x\, \sin y\, (3d\phi_1-\cos 2z\,\, d\phi_2)\Big)\,,\nn\\
P_4 &=&  \sin 2x\, \sin y\, dz +
      \ft34 \cos^2 y\, d\phi_1 + 
    \ft18(3-\cos 2y)\, \cos 2z\,\, d\phi_2 \,,\nn\\
P_5 &=& \fft{\sqrt3}{8}\, \Big((1-3\cos2y)\,  d\phi_1 + 2\cos^2 y\, 
          \cos2z\,\, d\phi_2\Big)\,.
\eea
%%%%%%
With the scaling chosen here, the metric is Einstein with $R_{ab} =
  6 g_{ab}$.

    The $SU(3)_{\rm maximal}$ connection is given by
%%%%%
\bea
Q_1 &=& -\cos 2x\, \sin y\, dz + \ft34 \sin 2x\, \cos^2 y\, d\phi_1
  + \ft18 (3 -\cos 2y\, )\sin 2x\, \cos 2z\,\, d\phi_2\,,\nn\\
Q_2 &=& - \sin x\, \cos y\, dz +\ft34 \cos x\, \sin 2y\,\, d\phi_1 -
  \ft14 \cos x\, \sin 2y\, \cos 2z\,\, d\phi_2\,,\nn\\
Q_3 &=& -\sin x\, dy - \ft12 \cos x\, \cos y\, \sin 2z\,\, d\phi_2\,.
\eea
%%%%%

\section{Einstein Metrics on the $SO(5)$ Group Manifold}

   Let $L_{AB}$ be the left-invariant 1-forms of $SO(5)$.  They are
antisymmetric, $L_{AB}=-L_{BA}$, with $1\le A\le5$ and $1\le B\le 5$,
and they satisfy
%%%%%
\be
  dL_{AB} = L_{AC}\wedge L_{CB}\,.
\ee
%%%%%
It is sometimes convenient to define
%%%%%
\be
\sigma_i = L_{1i}\,,\qquad \td\sigma_i = L_{2i}\,,\qquad \nu=L_{12}\,,\qquad
\hbox{where}\ \ 3\le i\le 5\,,\quad 3\le j\le 5\,.
\ee
%%%%%

    We find a total of 4 inequivalent Einstein metrics on $SO(5)$.  We 
obtain these by considering two different classes of metric, associated with
two different embeddings of an $SO(3)$ subgroup in $SO(5)$.  The first
class yields 3 inequivalent Einstein metrics:

\subsection{The $SO(3)_{\rm canonical}$ class}

   For this class, we make a decomposition in which the subgroup
$SO(3)_{\rm canonical} \subset SO(4)\subset SO(5)$ subgroup, generated by 
$L_{ij}$,  is manifest:
%%%%%
\be
ds_{10}^2 = x_1\, \sigma_i^2 + x_2\, \td\sigma_i^2 + x_3\, (L_{34}^2 
  + L_{35}^2 + L_{45}^2) + x_4 \, \nu^2\,.\label{so5can}
\ee
%%%%%
We may take the magnitude $V$ of the volume form to be defined by 
$\prod_a e^a = V\, \prod_{A<B} L_{AB}$, and so
%%%%%
\be 
V= (x_1 x_2 x_3)^{3/2}\, x_4^{1/2}\,.
\ee
%%%%%

 We obtain 3 inequivalent Einstein metrics as follows, with
the first being the standard bi-invariant metric:
\begin{itemize}

\item[(1)] Metric I:
%%%%%
\be
(x_1,x_2,x_3,x_4) = (1,1,1,1)\,,\qquad \lambda =\fft{3}{2}\,,
\qquad I_1= \fft{243}{32}\,,\qquad
                                   I_2 = 10\,.\label{so5met1}
\ee
%%%%%

\item[(2)] Metric II:
%%%%%
\be
(x_1,x_2,x_3,x_4) = \Big(\ft{7}{2},  \ft{7}{2} , 1,
                         \ft{ 19}{4} \Big)\,, \qquad\lambda=\fft{57}{98}\,,
                   \qquad
                 I_1 = \fft{ 601692057 \sqrt{19} }{421654016}\,,\qquad
         I_2 = \fft{240}{19}\,.\label{so5met2}
\ee
%%%%%

\item[(3)] Metric III:
%%%%%
\be
(x_1,x_2,x_3,x_4)= (1,2, 1, 2)\,,\qquad \lambda=\fft{9}{8}\,, 
     \qquad I_1= \fft{59049}{8192}\,,
\qquad I_2 =\fft{98}{9}\,.\label{so5met3}
\ee
%%%%%

\end{itemize}

\subsection{The $SO(3)_{\rm maximal}$ class}

   We can obtain a fourth inequivalent Einstein metric by choosing a 
basis for the $SO(5)$ left-invariant 1-forms in which the maximal
$SO(3)$ subgroup of $SO(5)$ is made manifest.  This subgroup is generated
by
%%%%%
\bea
Z_8 &=& \sqrt{\fft{6}{5}}\, \Big(L_{35} + \fft1{\sqrt3}\,(L_{13}+L_{24})\Big)\,,
\qquad
Z_9= \sqrt{\fft{6}{5}}\, \Big(L_{45} + \fft1{\sqrt3}\,(L_{23}-L_{14})\Big)
  \,,\nn\\
Z_{10} &=& \sqrt{\fft{2}{5}}\Big( 2 L_{12} + L_{34}\Big)\,.\label{Zso3}
\eea
%%%%%
The remaining generators are
%%%%%
\bea
Z_1 &=& \fft{2}{\sqrt5}\Big( L_{35} - \fft{\sqrt3}{2} (L_{13} + L_{24})
  \Big)\,,\qquad Z_2 = L_{13} - L_{24}\,,\nn\\
Z_3 &=& \fft{2}{\sqrt5}\Big( L_{45} - \fft{\sqrt3}{2} (L_{23} - L_{14})
  \Big)\,,\qquad Z_4 = L_{23} + L_{14}\,,\nn\\
Z_5 &=& \sqrt{\fft{2}{5}}\, (L_{12} - 2 L_{34})\,, \qquad
  Z_6= \sqrt2\, L_{15}\,,\qquad  Z_7= \sqrt2\, L_{25}\,.\label{Zs}
\eea
%%%%%

   In this basis, we consider the class of $SO(5)$ metrics
%%%%%
\be
ds_{10}^2 = y_1\, (Z_1^2+ Z_2^2 +Z_3^2 + Z_4^2) + y_2\, Z_5^2 +
  y_3\, (Z_6^2 + Z_7^2) + y_4\, (Z_8^2 + Z_9^2 + Z_{10}^2)\,.
\ee
%%%%%%
As well as the ``round'' Einstein metric $y_1=y_2=y_3=y_4$, which repeats
(\ref{so5met1}) above, we obtain a new Einstein metric:
\begin{itemize}

\item[(4)] Metric IV:
%%%%%
\be
(y_1,y_2,y_3,y_4) = (26,26,26,1)\,,\quad \lambda = \fft{69}{1352}\,,\quad
I_1 = \fft{1564031349}{308915776\sqrt{26}}\,,\quad
I_2= \fft{16705}{1058}\,.\label{so5met4}
\ee
%%%%%
Note that here, we again define $V$ via $\prod_a e^a = V\, 
\prod_{A<B} L_{AB}$, and so in this case we have
%%%%%
\be
V= 32 y_1^2 \, y_2^{1/2}\, y_3\, y_4^{3/2}\,.
\ee
%%%%%

\end{itemize}

    Since each of the 4 Einstein metrics (\ref{so5met1}), (\ref{so5met2}),
(\ref{so5met3}) and (\ref{so5met4}) has different values for the invariants
$I_1$ and $I_2$, they are definitely all inequivalent. Included among them
are the standard bi-invariant metric (\ref{so5met1}) and the second 
Einstein metric (\ref{so5met4}) whose
existence was established in \cite{Je2}.  The remaining two Einstein metrics
(\ref{so5met2}) and (\ref{so5met3}) appear to be new.

\section{Einstein Metrics on Cosets $SO(5)/H$}

\subsection{Einstein metrics on $SO(5)/U(1)$}

   Here, we choose the $SO(5)$ basis
%%%%%
\bea
X_1 &=& L_{13} + L_{24}\,,\qquad X_2= L_{23}- L_{14}\,,\qquad
  X_3= L_{13} - L_{24}\,,\qquad X_4= L_{23} + L_{14}\,,\nn\\
X_5 &=& \sqrt2\, L_{15}\,,\qquad X_6=\sqrt2\, L_{25}\,,\qquad
  X_7= \sqrt2\, L_{35}\,,\qquad X_8= \sqrt2\, L_{45}\,,\nn\\
X_9 &=& c\, L_{12} + s\, L_{34}\,,\qquad X_{10}= c\, L_{34} - s\, L_{12}\,,
\label{9basis}
\eea
%%%%%
where $c=\cos\theta$, $s=\sin\theta$.  The two commuting generators are taken
to be $X_9$ and $X_{10}$.  The angle $\theta$ parameterises the embedding of 
the $U(1)$ denominator group in the maximal torus $T^2$. 
The cosets $SO(5)/U(1)$ are obtained by dividing
out by $X_{10}$, and writing the coset metric as
%%%%%
\be
ds_9^2= z_1\, (X_1^2+X_2^2) + z_2\, (X_3^2+X_4^2) +
   z_3\, (X_5^2+X_6^2) + z_4\, (X_7^2 + X_8^2) + z_5\, X_9^2\,.
\ee
%%%%%

  It appears that the non-trivial range for $\theta$ is $0\le\theta\le \pi/4$.
Angles outside this range give metrics equivalent to ones with $\theta$ 
inside the range.  The actual allowed values of $\theta$ for which the
local metrics can be extended smoothly onto complete manifolds will form a
discrete but infinite set within the range, characterised by
coprime integers $(k,\ell)$ defining rational numbers $k/\ell$.  

   Solving the Einstein conditions, we find the following:
%%%%%
\bea
\theta=0:&& \ \ \hbox{\bf 3 Inequivalent Einstein Metrics:}\nn\\
(z_1,z_2,z_3,z_4,z_5)&=& (1,1,\fft{3-\sqrt5}{4},1,\fft{3-\sqrt5}{2})\,,\qquad
(1,1,\fft{3+\sqrt5}{4},1,\fft{3+\sqrt5}{2})\,,\nn\\
&&(1,1.8522,1.5490,0.9614,3.2786)\nn\\
&&\nn\\
\theta=\fft{\pi}{4}:&& \ \ \hbox{\bf 1 Inequivalent Einstein Metric}\nn\\
(z_1,z_2,z_3,z_4,z_5) &=& (1, 1.1841,0.2244,1.0924,0.8504)\nn\\
&&\nn\\
0<\theta <\fft{\pi}{4}:&& \ \ \hbox{\bf 
  2 Inequivalent Einstein Metrics; Example
for $\theta=\fft{\pi}{6}$:} \nn\\
(z_1,z_2,z_3,z_4,z_5) &=& (1,1.4222,1.2123,0.2888,1.8813)\,,\quad
   (1,1.0950, 0.2049,1.0476,0.5375)\nn
\eea
%%%%%
Note that for $\theta=0$ we actually obtain 4 solutions for the coefficients,
namely the three listed plus a fourth (numerical) solution.  This is 
equivalent, up to permutation of generators, to the third listed solution.
For $\theta=\pi/4$, we actually obtain two (numerical) solutions, but the
second is equivalent to the one listed.  For generic $\theta$, \ie in the
range $0<\theta<\pi/4$, we obtain exactly two (numerical) solutions,
and they are inequivalent.  According to \cite{bohker}, only one Einstein
metric was known previously for each $\theta$.

\subsection{Einstein Metrics on $SO(5)/T^2$}

   To construct these, we begin with the $SO(5)$ basis $X_a$ defined in
equation (\ref{9basis}), and then omit the generators associated with
the Cartan subalgebra $X_9$ and $X_{10}$.  Thus we consider
%%%%%
\be
ds_8^2 = w_1\, (X_1^2+X_2^2) + w_2\, (X_3^2+X_4^2) +
  w_3\, (X_5^2+X_6^2) + w_4\, (X_7^2+X_8^2)\,.
\ee
%%%%%
We obtain two solutions with
%%%%%
\crampest{
\be
(w_1,w_2,w_3,w_4) = (\fft{24-4\sqrt6}{15}, \fft{24-4\sqrt6}{15} , 
     \fft{7-2\sqrt6}{5}, 1)\,,\quad
(\fft{24+4\sqrt6}{15}, \fft{24+4\sqrt6}{15} ,
     \fft{7+2\sqrt6}{5}, 1)\,,\label{two8}
\ee}
%%%%%
and four solutions with
%%%%%
\be
(w_1,w_2,w_3,w_4)= (4,2,3,1)\,,\quad (\fft{2}{3}, \fft{4}{3},\fft1{3},1)
\,,\quad (2,4,3,1)\,,\quad (\fft{4}{3}, \fft{2}{3}, \fft1{3}, 1)\,.
\label{four8}
\ee
%%%%%
In fact the two solutions in (\ref{two8}) are equivalent up to permutation
and scaling.  Similarly, the four solutions in (\ref{four8}) are equivalent
up to permutation and scaling.  This might be suspected from the values
of the invariants
%%%%%
\be 
I_1 = \lambda^4\, V\,,\qquad I_2= |\hbox{Riem}|^2\, \lambda^{-2}\,,
\ee
%%%%%
which are given by $(I_1,I_2)=(40/27, 1449/100)$ for both of the solutions
in (\ref{two8}), and by $(I_1,I_2)=(3/2,16)$ for all four of the
solutions in (\ref{four8}).  Explicit calculations show that indeed the
pair (\ref{two8}) can be related by relabelling and scaling, as can the
quartet (\ref{four8}).

   Thus we have in total two inequivalent 
Einstein metrics on $SO(5)/T^2$.  One
Einstein metric corresponds to taking either of the equivalent pair
in (\ref{two8}).  The other corresponds to taking any one of the
equivalent quadruplet in (\ref{four8}).\footnote{Reference \cite{bohker}
states that three inequivalent Einstein metrics on $SO(5)/T^2$ are known.
The result is attributed to Sakane \cite{sakane}.  In fact Sakane
obtained the two solutions (\ref{two8}) and the four solutions (\ref{four8}),
but did not explicitly discuss equivalences among them.  We suspect that
the two equivalent solutions (\ref{two8}) were mistakenly counted as being
distinct.}

\subsection{Seven-Dimensional Einstein Spaces $SO(5)/SO(3)$}

\subsubsection{Einstein metrics on $SO(5)/SO(3)_{\rm canonical}$}

    Here, we take the $SO(3)$ subgroup to be generated by the subset
$L_{ij}$, where $3\le i\le 5$ and $3\le j\le 5$.  The metric on the
coset $SO(5)/SO(3)_{\rm canonical}$, which is the Stiefel manifold
$V_{5,2}$, is then taken to be of the form
%%%%%
\be
ds_7^2 = u_1\, \sigma_i^2 + u_2\, \td\sigma_i^2 + u_3\, \nu^2\,,
\ee
%%%%%
where as before, $\nu=L_{12}$, $\sigma_i=L_{1i}$ and $\td\sigma_i=L_{2i}$,
where $3\le i\le 5$.  We can obtain one Einstein metric in this class,
by taking
%%%%%
\be 
(u_1,u_2,u_3)= (1,1,\fft{3}{2})\,.
\ee
%%%%%
It satisfies the Einstein equations $R_{ab}=\lambda\, g_{ab}$ with
%%%%%
\be
\lambda= \fft{9}{4}\,.
\ee
%%%%%

   This metric admits two Killing spinors \cite{carowa}, satisfying 
%%%%%
\be
  \nabla_a\eta - \fft{\im}{2}\,m\,  \Gamma_a\eta =0\,,
\ee
%%%%%
where $6m^2= \lambda= 9/4$.

\subsubsection{Einstein metrics on $SO(5)/SO(3)_{\rm maximal}$}

   In this case, we take the $SO(3)$ subgroup to be maximal in $SO(5)$.
Under this embedding, we have the group decompositions
%%%%%
\be
{\bf 4} \longrightarrow {\bf 4}\,,\qquad
{\bf 5}\longrightarrow {\bf 5}\,,\qquad
{\bf 10}\longrightarrow {\bf 7} + {\bf 3}\,.
\ee
%%%%%
The subgroup $SO(3)_{\rm maximal}$ is generated by $(Z_8,Z_9,Z_{10})$
defined in (\ref{Zso3}).  The remaining coset generators are $Z_a$ for
$1\le a\le 7$, as defined in (\ref{Zs}).  

   There is a unique metric (up to scaling) on 
coset $SO(5)/SO(3)_{\rm maximal}$, given by
%%%%%
\be
ds_7^2 = Z_1^2+Z_2^2+Z_3^2+Z_4^2 + Z_5^2 + 
Z_6^2+Z_7^2\,.
\ee
%%%%%
It is Einstein, satisfying $R_{ab}=\lambda\, g_{ab}$ with
%%%%%
\be
\lambda= \fft{27}{20}\,.
\ee
%%%%%

   This metric admits one Killing spinor \cite{carowa}.

\subsubsection{Einstein metrics on $SO(5)/SO(3)_L$}

   Here, we take the subgroup $SO(3)_L$ in the isomorphism
$SO(4) = SO(3)_L\times SO(3)_R$ as the denominator in the coset. The
coset has the topology $S^7$.  Taking $1\le a\le 4$, we split the
$SO(5)$ generators as $L_{a5}$ and $L_{ab}$, and the decompose $L_{ab}$
into their self-dual and anti-self-dual parts:
%%%%%
\bea
L_1&=& \fft1{\sqrt2}\, (L_{12}-L_{34})\,,\qquad
L_2= \fft1{\sqrt2}\, (L_{23}-L_{14})\,,\qquad
L_3=\fft1{\sqrt2}\, (L_{31}-L_{24})\,,\nn\\
R_1&=& \fft1{\sqrt2}\, (L_{12}+L_{34})\,,\qquad
R_2= \fft1{\sqrt2}\, (L_{23}+L_{14})\,,\qquad
R_3=\fft1{\sqrt2}\, (L_{31}+L_{24})\,.
\eea
%%%%%
We then consider metrics
%%%%%
\be
ds_7^2 = v_1\, (L_{15}^2 + L_{25}^2 + L_{35}^2 + L_{45}^2) +
        v_2\, (R_1^2 + R_2^2 + R_3^2)\,.
\ee
%%%%%
We obtain two inequivalent Einstein metrics, with
%%%%%
\be
(v_1,v_2) = (1,2)\,, \qquad  (1,\fft{2}{5})\,.
\ee
%%%%%
These correspond to the round $S^7$, and the squashed $S^7$ of
Jensen \cite{jen},
respectively.  They satisfy $R_{ab}=\lambda \, g_{ab}$ with $\lambda= 3/2$
and $\lambda= 27/10$ respectively.

   The round $S^7$ admits 8 Killing spinors, while the squashed Einstein
metric admits 1 Killing spinor \cite{awdupo}.

\section{Einstein Metrics on $G_2$}

   The exceptional group $G_2$ is a subgroup of $SO(7)$.  Let the
generators for $SO(7)$ be $T_{AB} =- T_{BA}$. If we decompose 
the $SO(7)$ fundamental index $A$ as $A=(i,\hat i, 7)$, where $i=1,2,3$, 
$\hat i=\hat 1,\hat 2,\hat 3= 4,5,6$, then the $14=3+3+8$
 generators of $G_2$ can be taken to be \cite{awdupo}
%%%%%
\bea
G_i &=& T_{i7} + \ft12 \ep_{ijk}\, T_{\hat j\hat k}\,,\qquad
G_{ij} = T_{ij} + T_{\hat i \hat j}\,,\nn\\
G_{i\hat j} &=& \fft{2}{\sqrt3}\Big(-T_{i\hat j} -\ft12 T_{j\hat i} + 
  \ft12 \delta_{ij}\, T_{k\hat k} -\ft12 \ep_{ijk}\, T_{\hat k 7}\Big)\,.
\label{G2subalg}
\eea
%%%%%
Note that $G_{i\hat j}$ is traceless; $G_{i\hat i}=0$.

 If we associate left-invariant 1-forms $\sigma_a=\{\sigma_i,
\sigma_{ij},\sigma_{i\hat j}\}$ with each $G_2$ generator,
then we may write
%%%%%
\bea
\sigma_i &\cong& L_{i7} + \ft12 \ep_{ijk}\, L_{\hat j\hat k}\,,\qquad
\sigma_{ij} \cong L_{ij} + L_{\hat i \hat j}\,,\nn\\
\sigma_{i\hat j} &\cong& \fft{2}{\sqrt3}\Big(-L_{i\hat j} -\ft12 L_{j\hat i} +
  \ft12 \delta_{ij}\, L_{k\hat k} -\ft12 \ep_{ijk}\, L_{\hat k 7}\Big)\,,
\eea
%%%%%
where $L_{AB}=-L_{BA}$ are left-invariant 1-forms for $SO(7)$, satisfying
$dL_{AB}=L_{AC}\wedge L_{CB}$.  When evaluating the exterior derivatives of
$\sigma_a$, we then project into
the subspace of 2-forms spanned by wedge products of the $\sigma_a$, in order
to read off the Cartan-Maurer equations for the left-invariant 1-forms 
of $G_2$.\footnote{This projection procedure is the implementation, 
at the level of the exterior algebra of the 1-forms, of the fact that
the commutators of the $G_2$ generators $G_i$, $G_{ij}$ and $G_{i\hat j}$ 
defined by (\ref{G2subalg}) close on themselves.}

   We can consider the class of left-invariant metrics of the form
%%%%%
\be
ds_{14}^2 = \sum_{a=1}^6 x_a\, E^+_a\, E^-_a + x_7\, H_1^2 + x_8\, H_2^2\,,
\label{G2met1}
\ee
%%%%%
where $E_a^+$ denotes the six left-invariant 1-forms corresponding to
the six positive roots of $G_2$, and correspondingly, $E^-_a$ denotes
the six left-invariant conjugate 1-forms for the negative roots.  $H_1$ and
$H_2$ denote the left-invariant 1-forms for the two Cartan generators.

  In terms of the left-invariant 1-forms $\sigma_i$, $\sigma_{ij}$ and
$\sigma_{i\hat j}$ defined above, we have
%%%%%
\bea
E_1^+ &=& \sigma_{3\hat 2} - \sigma_{2\hat 3}+ 
           \fft{\im}{\sqrt3}\,(\sigma_{23} -2 \sigma_1)\,,\nn\\
E_2^+ &=& \sigma_{3\hat1}-\sigma_{1\hat3} +
   \fft{\im}{\sqrt3}\, (\sigma_{31}-2\sigma_2)\,,\nn\\
E_3^+ &=& \sigma_{2\hat1}-\sigma_{1\hat2} -
          \fft{\im}{\sqrt3}\, (\sigma_{12} - 2\sigma_3)\,,\nn\\
E_4^+ &=& \sigma_{1\hat2}+\sigma_{2\hat1} -\im\, \sqrt3\, \sigma_{12}\,,\nn\\
E_5^+ &=& \sigma_{1\hat3}+\sigma_{3\hat1} + \im\, \sqrt3\, \sigma_{31}\,,\nn\\
E_6^+ &=& \sigma_{2\hat3}+\sigma_{3\hat2} + \im\, \sqrt3\, \sigma_{23}\,,
\label{G2E}
\eea
%%%%%
and
%%%%%
\be
H_1= \sigma_{1\hat1} -\sigma_{2\hat2}\,,\qquad 
     H_2= -\sqrt3\, (\sigma_{1\hat1}+\sigma_{2\hat2})\,.\label{G2H}
\ee
%%%%%
The left-invariant 1-forms $E^-_a$ corresponding to the negative
roots are obtained from the $E^+_a$ in (\ref{G2E}) by reversing the
sign of $\im$. 
The weights of the $E_a^+$ under $(H_1,H_2)$ are 
%%%%%
\be
(1,1)\,,\quad (1,-1)\,,\quad (0,2)\,,\quad (2,0)\,,\quad (1,3)\,,\quad
(1,-3)\,,
\ee
%%%%%
for $a=1$ up to $a=6$ respectively.

\subsection{The $SU(2)_{\rm diag}$ class}

   The metric (\ref{G2met1}) can thus be written as 
%%%%%
\bea
ds_{14}^2 &=& x_1\, \Big[ (\sigma_{2\hat 3} -\sigma_{3\hat 2})^2 + 
     \ft13 (\sigma_{23} -2 \sigma_1)^2\Big] +
 x_2\, \Big[ (\sigma_{3\hat 1} -\sigma_{1\hat 3})^2 +
     \ft13 (\sigma_{31} -2 \sigma_2)^2\Big] \nn\\
&&+
 x_3\, \Big[ (\sigma_{1\hat 2} -\sigma_{2\hat 1})^2 +
     \ft13 (\sigma_{12} -2 \sigma_3)^2\Big]\nn\\
&& + x_4\, \Big[3 \sigma_{23}^2 + (\sigma_{2\hat3}+\sigma_{3\hat2})^2\Big]
+x_5\, \Big[3 \sigma_{31}^2 + (\sigma_{3\hat1}+\sigma_{1\hat3})^2\Big]+
x_6\, \Big[3 \sigma_{12}^2 + (\sigma_{1\hat2}+\sigma_{2\hat1})^2\Big]\nn\\
&& + x_7\, (\sigma_{1\hat1} -\sigma_{2\hat2})^2 +
     3 x_8\, (\sigma_{1\hat1} + \sigma_{2\hat2})^2\,.\label{G2met12}
\eea
%%%%%
As will be seen in section \ref{su2diagsec} below, the basis used here is
naturally adapted to the embedding of the $SU(2)_{\rm diag}$ subgroup
in $G_2$, where $SU(2)_{\rm diag}$ is the diagonal $SU(2)$ in the
$SU(2)\times SU(2)$ subgroup of $G_2$.

   We find two choices (up to overall scaling) for the coefficients $x_i$
that yield Einstein metrics, namely
%%%%%
\bea
(x_1,\ldots, x_8)=(3,3,3,1,1,1,1,1):&& \lambda=\fft{1}{3}\,,\qquad 
   I_1=\fft1{81}\,,\qquad I_2= 14\,,\label{2g2met}\\
(x_1,\ldots, x_8)=(\ft{11}{3},\ft{11}{3},\ft{11}{3}, 1,1,1,
1,1):&&\lambda=\fft{37}{121}\,,
\qquad I_1=\fft{(37)^7}{27\cdot (11)^{11}}\,,\qquad
        I_2= \fft{19346}{1369}\,.\nn
\eea
%%%%%
Here we take the volume to be
$V=x_1\, x_2\, 
x_3\, x_4\, x_5\, x_6\, \sqrt{x_7\, x_8}$ when calculating the invariant
$I_1$. Note that the first metric in (\ref{2g2met}) is the
bi-invariant one.  The second is the non-bi-invariant metric obtained
in the analysis of D'Atri and Ziller \cite{datzil}.

\subsection{The $SU(2)\times SU(2)$ class}

    We can obtain further Einstein metrics on $G_2$ by considering a
different choice of basis for the metric, adapted this time to the
$SU(2)\times SU(2)$ subgroup of $G_2$ (see sections \ref{su2Lsec} and
\ref{su2Rsec} below).  If we take
%%%%%
\bea
ds_{14}^2 &=& y_1\, [(\sigma_{1\hat 2} -\sigma_{2\hat 1})^2 +
 (\sigma_{1\hat 3} -\sigma_{3\hat 1})^2 +
     (\sigma_{2\hat 3} -\sigma_{3\hat 2})^2]\nn\\
&& +
 y_2\,  [(\sigma_{1\hat 2} + \sigma_{2\hat 1})^2 +
 (\sigma_{1\hat 3} + \sigma_{3\hat 1})^2 +
     (\sigma_{2\hat 3} + \sigma_{3\hat 2})^2]\nn\\
&& + y_3\, [(\sigma_{23}-\sigma_1)^2 + (\sigma_{31}-\sigma_2)^2 +
            (\sigma_{12}-\sigma_3)^2] \nn\\
&&+
y_4\, [(\sigma_{23}+\sigma_1)^2 + (\sigma_{31}+\sigma_2)^2 +
            (\sigma_{12}+\sigma_3)^2] \nn\\
&& + y_5\, (\sigma_{1\hat1} - \sigma_{2\hat2})^2 + 
     y_6\, (\sigma_{1\hat1} + \sigma_{2\hat2})^2\,,\label{G2more}
\eea
%%%%%
then the metric is Einstein if
%%%%%
\be
y_1=3\,,\quad y_2=1\,,\quad y_4= \fft{(7 y_3-6)(6-y_3)}{15 y_3}\,,\quad
   y_5=1\,,\quad y_6=3\,,\label{g2ysol}
\ee
%%%%%
and $y_3$ is a root of the quartic polynomial
%%%%%
\be
(y_3-3)(35 y_3^3 - 303 y_3^2 + 666 y_3-378)=0\,.\label{ypol}
\ee
%%%%%
The root $y_3=3$ reproduces the first Einstein metric listed in 
(\ref{2g2met}).   The three roots of cubic polynomial factor
in (\ref{ypol}) are given by
%%%%%
\be
y_3= \fft{101}{35} + \fft{2\sqrt{2431}}{35}\, 
\cos\Big(\fft{\theta+ 2\pi n}{3}\Big)\,,
\qquad \hbox{for}\qquad n=0, 1, 2\,,\label{y3root}
\ee
%%%%%
where
%%%%%
\be
\cos\theta= \fft{84671}{(2431)^{3/2}}\,.
\ee
%%%%%
These roots are all real and positive,
and furthermore $y_4$, given
in (\ref{g2ysol}), is positive in all these cases.  This yields three
further Einstein metrics on the $G_2$ group manifold, which are all
inequivalent, and they are all inequivalent to the two already listed in
(\ref{2g2met}).  
The Einstein constant and the 
invariants $I_1= \lambda^7\, V$ and $I_2=\lambda^{-2}\, |\hbox{Riem}|^2$
for the three additional Einstein metrics, for $n=0$, $n=1$ and $n=2$ 
in (\ref{y3root}), are given numerically by
%%%%%
\bea
n=0:&&\lambda\approx 0.40067\,,\qquad
       I_1 \approx 0.021017\,,\qquad I_2\approx 20.84408\,,\nn\\
n=1:&& \lambda\approx 0.60962\,,\qquad
  I_1\approx 0.012100 \,,\qquad I_2\approx 19.35457\,,\nn\\
n=2:&& \lambda\approx 0.35162\,,\qquad
    I_1\approx 0.036879 \,,\qquad I_2\approx 14.30375\, \label{G2345}
\eea
%%%%%
For comparison, the numerical values of the invariants for 
the two Einstein metrics listed in
(\ref{2g2met}) are $(I_1,I_2)\approx (0.037037,14)$ and 
$(I_1,I_2)\approx (0.036970,14.13148)$ respectively. 

\subsection{The $SU(2)_{\rm max}$ class}

  There is a third choice of basis that enables us to find one further
inequivalent Einstein metric on $G_2$.  This basis is adapted to the
maximal $SU(2)$ subgroup in $G_2$ (see section \ref{su2maxsec} below),
and in it the metric is given by
%%%%%
\bea
ds_{14}^2 &=& x_1\,\Big[\ft13  (\sigma_{23}- 2 \sigma_1)^2+
   ( \sigma_{3\hat 2} - \sigma_{2\hat 3})^2\Big] +
x_2\, \Big[\ft13(\sigma_{31}-2\sigma_2)^2+ (\sigma_{3\hat1}-\sigma_{1\hat3})^2
\Big]\,,\nn\\
&&+ x_3\, \Big[3 \sigma_{12}^2 +  (\sigma_{1\hat2}+\sigma_{2\hat1})^2\Big]
 + x_4\, \Big[3 \sigma_{31}^2 +(\sigma_{1\hat3}+\sigma_{3\hat1})^2\Big]
\,,\\
&& +x_5 \Big[ \ft13 (\ft{3\sqrt3}{\sqrt5}\, \sigma_{23} -
                      2\sigma_3 + \sigma_{12})^2 +
  (\ft{\sqrt3}{\sqrt5}\, (\sigma_{2\hat3}+\sigma_{3\hat2})  +
            \sigma_{1\hat2}-\sigma_{2\hat1})^2 \Big] +
  \ft43  x_6\, (2\sigma_{1\hat1}+3\sigma_{2\hat2})^2\nn\\
&&+\ft1{96} x_7 (\sqrt{15}\, \sigma_{23} + 6 \sigma_3 + \sigma_{12})^2 +
\ft1{288} x_7 (\sqrt{15}(\sigma_{2\hat3}+\sigma_{3\hat2}) -
         9(\sigma_{1\hat2}+\sigma_{2\hat1}))^2 +
\ft1{588} x_7 (4\sigma_{1\hat1}-\sigma_{2\hat2})^2\,.\nn
\eea
%%%%%
We find two choices of coefficients (up to scale) that give Einstein
metrics, namely
%%%%%
\bea
(x_1,x_2,x_3,x_4,x_5,x_6,x_7) &=& (1,1,\ft13,\ft13,\ft5{14},\ft3{28},
  28)\,,\label{binvagain}\\
(x_1,x_2,x_3,x_4,x_5,x_6,x_7) &=& (1,1,\ft13,\ft13,\ft5{14},\ft3{28},
\ft{28}{85})\,.\label{G26}
\eea
%%%%%
The first case is just the standard bi-invariant metric but (\ref{G26})
is new, with
%%%%%
\be
\lambda= \fft{26}{17}\,,\qquad I_1=\fft{(26)^7}{3^4\, 17^8\, 5\sqrt{85}}
\,,\qquad I_2=\fft{5719}{260}\,.
\ee
%%%%
The invariants $I_1$ and $I_2$ are different from those we found in the 
previous five Einstein metrics.

   To summarise, we have found six inequivalent Einstein metrics on the
$G_2$ group manifold, of which the last four, given by (\ref{y3root}) and
(\ref{G26}), appear to be new.  

\section{Einstein Metrics on Cosets $G_2/H$}

\subsection{$G_2/(SU(2)\times SU(2))$}\label{G2so4sec}

   The $SU(2)\times SU(2)$ subgroup is generated by
%%%%%
\be
X_i= G_i -\ft12 \ep_{ijk}\, G_{jk}\,,\qquad 
Y_i= G_i + \ft12 \ep_{ijk}\, G_{jk}\,,
\ee
%%%%
where $X_i$ are the generators of one $SU(2)$ factor, and $Y_i$ generates 
the other.  The space is isotropy-irreducible, and so there is 
just one Einstein metric \cite{wolf}.  It is given by
%%%%%
\bea
ds_8^2 &=& 3[(\sigma_{2\hat 3}- \sigma_{3\hat2})^2 + 
             (\sigma_{3\hat 1}- \sigma_{1\hat3})^2 +
             (\sigma_{1\hat 2}- \sigma_{2\hat1})^2] \nn\\
&&+(\sigma_{2\hat 3}+ \sigma_{3\hat2})^2 +
             (\sigma_{3\hat 1}+ \sigma_{1\hat3})^2 +
             (\sigma_{1\hat 2}+ \sigma_{2\hat1})^2\nn\\
&& + 3 (\sigma_{1\hat1} + \sigma_{2\hat2})^2 +
          (\sigma_{1\hat 1}-\sigma_{2\hat2})^2  \,.
\eea
%%%%%
It satisfies $R_{ab} =\ft23 g_{ab}$.

\subsection{$G_2/SU(2)_{\rm diag}$}\label{su2diagsec}

   We can obtain an Einstein metric on the 11-dimensional coset space
$G_2/SU(2)_{\rm diag}$, where $SU(2)_{\rm diag}$ is the diagonal 
$SU(2)$ subgroup in $SU(2)\times SU(2)$.  It is therefore
associated
with $\sigma_{12}$, $\sigma_{23}$ and $\sigma_{31}$.  We can then obtain
$G_2$ invariant metrics on the coset, with        
%%%%%
\bea
ds_{11}^2 &=& y_1\, [(\sigma_{2\hat 3} -\sigma_{3\hat 2})^2 +
    (\sigma_{3\hat 1} -\sigma_{1\hat 3})^2 +
     (\sigma_{1\hat 2} -\sigma_{2\hat 1})^2] \nn\\
&&+ \ft13 y_2\, [(\sigma_{23} -2 \sigma_1)^2 +
                  (\sigma_{31} -2 \sigma_2)^2 +(\sigma_{12} -2 \sigma_3)^2]
        \\
&& + y_3\, [(\sigma_{2\hat3}+\sigma_{3\hat2})^2 + 
       (\sigma_{3\hat1}+\sigma_{1\hat3})^2 +
        (\sigma_{1\hat2}+\sigma_{2\hat1})^2
   + (\sigma_{1\hat1} -\sigma_{2\hat2})^2 +
          3 (\sigma_{1\hat1} + \sigma_{2\hat2})^2] \,.\nn
\eea
%%%%%
We then find that there is a Riemannian Einstein metric if
%%%%%
\be
y_2=\fft{y_1(27-5 y_1)}{9 + 5 y_1}\,,\qquad y_3=1\,,\label{y2sol}
\ee
%%%%%
where $y_1$ is the real, positive root of the quartic polynomial
%%%%%
\be
125 y_1^4 - 500 y_1^3 + 213 y_1^2 + 378 y_1 -972=0\,.\label{y1poly}
\ee
%%%%%
There is also a pseudo-Riemannian Einstein metric when $y_1$ is the real, 
negative root of (\ref{y1poly}).  Since $y_2$, given by (\ref{y2sol}), is
then also negative, the metric signature is $(6,5)$.

   Since the adjoint of $G_2$ decomposes under the $SU(2)\times SU(2)$
maximal subgroup as 
%%%%%
\be
{\bf 14} \longrightarrow (\ft32,\ft12)\quad \oplus\quad (1,0) 
\quad\oplus \quad (0,1)\,,
\label{su2su2}
\ee
%%%%
where we denote an $SU(2)$ representation by its spin $j$, it follows
that under $SU(2)_{\rm diag}$ we shall have
%%%%%
\be
{\bf 14} \longrightarrow (2) \quad \oplus \quad 3\times (1)\,.
\ee
%%%%%
(In other words, we have one spin-2 and three spin-1 representations in the
decomposition.)  One may define the Dynkin index of an $SU(2)$ embedding in
a group $G$ by
%%%%%
\be
I_D= \ft18 \sum_j \rho_j\,,\qquad  \rho_j= \ft23 j(j+1)(2j+1)\,,
\label{dynkin}
\ee
%%%%%
where the summation is taken over all the irreducible representations, labelled
by their spin $j$, in the decomposition of the adjoint of $G$.  Thus we see
that the Dynkin index for the $SU(2)_{\rm diag}$ subgroup in $G_2$ is 
given by
%%%%%
\be 
I_D= 4\,.
\ee
%%%%%
The Riemannian Einstein metric we have obtained here is therefore the one
listed as $G_2/SO(3)_4$ in \cite{bohker}, which was obtained in \cite{dicker}.

\subsection{$G_2/SU(2)_L$}\label{su2Lsec}

    Here, we consider the coset formed by dividing out by the $SU(2)_L$   
factor in the $SU(2)_L\times SU(2)_R$ subgroup described in section
\ref{G2so4sec}.  This amounts to factoring out the three terms proportional
to $y_4$ in (\ref{G2more}), which can be done provided the relations
%%%%%
\be
y_2=\ft13 y_1\,,\qquad y_5=\ft13 y_1\,,\qquad y_6=y_1
\ee
%%%%%
are imposed.   Metrics on the eleven-dimensional coset are therefore given
by
%%%%%
\bea
ds_{11}^2 &=& y_1\, [(\sigma_{1\hat 2} -\sigma_{2\hat 1})^2 +
 (\sigma_{1\hat 3} -\sigma_{3\hat 1})^2 +
     (\sigma_{2\hat 3} -\sigma_{3\hat 2})^2]\nn\\
&& +
 \ft13 y_1\,  [(\sigma_{1\hat 2} + \sigma_{2\hat 1})^2 +
 (\sigma_{1\hat 3} + \sigma_{3\hat 1})^2 +
     (\sigma_{2\hat 3} + \sigma_{3\hat 2})^2]\nn\\
&& + y_3\, [(\sigma_{23}-\sigma_1)^2 + (\sigma_{31}-\sigma_2)^2 +
            (\sigma_{12}-\sigma_3)^2] \nn\\
&& + \ft13 y_1\, (\sigma_{1\hat1} - \sigma_{2\hat2})^2 +
     y_1\, (\sigma_{1\hat1} + \sigma_{2\hat2})^2\,,
\eea
%%%%%
Imposing the Einstein condition $R_{ij}=\lambda g_{ij}$, we obtain two
solutions (up to overall scale):
%%%%%
\bea
(y_1,y_3)=(1,2):&& \lambda=\fft{5}{4}\,,\qquad I_1= 
 \fft{3125 \sqrt{5}}{512 \sqrt2}\,,\qquad I_2= \fft{257}{15}\,,\nn\\
(y_1,y_3)=(1,\ft27):&& \lambda=\fft{53}{28}\,,\qquad I_1=
\fft{(53)^{11/2}}{2^{19/2}\, 7^7}\,,\qquad
I_2= \fft{132517}{8427}\,.
\eea
%%%%%

    From (\ref{su2su2}) we see that under $SU(2)_L$, the adjoint of
$G_2$ decomposes as
%%%%%
\be
{\bf 14}\longrightarrow 2\times (\ft32) \quad\oplus\quad (1)
\quad \oplus \quad 3\times (0)\,,
\ee
%%%%%
and hence from (\ref{dynkin}) the Dynkin index of the $SU(2)_L$ embedding is
%%%%%
\be
I_D=3\,.
\ee
%%%%%
The two Einstein metrics we have obtained here are the ones denoted by
$G_2/SU(2)_3$ in \cite{bohker}, which were obtained in \cite{jen,Wa2}.

\subsection{$G_2/SU(2)_R$}\label{su2Rsec}

    Here, we consider the coset formed by dividing out by the $SU(2)_R$
factor in the $SU(2)_L\times SU(2)_R$ subgroup described in section
\ref{G2so4sec}.  This amounts to factoring out the three terms proportional
to $y_3$ in (\ref{G2more}), which can be done provided the relations
%%%%%
\be
y_2=\ft13 y_1\,,\qquad y_5=\ft13 y_1\,,\qquad y_6=y_1
\ee
%%%%%
are imposed.   Metrics on the eleven-dimensional coset are therefore given
by
%%%%%
\bea
ds_{11}^2 &=& y_1\, [(\sigma_{1\hat 2} -\sigma_{2\hat 1})^2 +
 (\sigma_{1\hat 3} -\sigma_{3\hat 1})^2 +
     (\sigma_{2\hat 3} -\sigma_{3\hat 2})^2]\nn\\
&& +
 \ft13 y_1\,  [(\sigma_{1\hat 2} + \sigma_{2\hat 1})^2 +
 (\sigma_{1\hat 3} + \sigma_{3\hat 1})^2 +
     (\sigma_{2\hat 3} + \sigma_{3\hat 2})^2]\nn\\
&& + y_4\, [(\sigma_{23}+\sigma_1)^2 + (\sigma_{31}+\sigma_2)^2 +
            (\sigma_{12}+\sigma_3)^2] \nn\\
&& + \ft13 y_1\, (\sigma_{1\hat1} - \sigma_{2\hat2})^2 +
     y_1\, (\sigma_{1\hat1} + \sigma_{2\hat2})^2\,,
\eea
%%%%%
Imposing the Einstein condition $R_{ij}=\lambda g_{ij}$, we obtain two
solutions (up to overall scale) with $y_1=0$ and 
$315 y_4^2-144 y_4 +4=0$:
%%%%%
\bea
(y_1,y_4)=(1,\fft{2(12+\sqrt{109})}{105}):&& 
\lambda=\fft{44-\sqrt{109}}{28}\,,\qquad I_1=
 \fft{(91-8\sqrt{109})^{11/2}}{1536\sqrt6(12-\sqrt{109})}
\,,\nn\\
&& I_2= \fft{ 14448791 - 425072 \sqrt{109}}{ 2818800}\,,\nn\\
&& \nn\\
(y_1,y_4)=(1,\fft{2(12-\sqrt{109})}{105}):&& \lambda=\fft{44+\sqrt{109}}{28}
\,,\qquad I_1=
 \fft{(91+8\sqrt{109})^{11/2}}{1536\sqrt6(12+\sqrt{109})}\,,\nn\\
&&I_2= \fft{ 14448791 + 425072 \sqrt{109}}{ 2818800}\,.
\eea
%%%%%

    From (\ref{su2su2}) we see that under $SU(2)_R$, the adjoint of
$G_2$ decomposes as
%%%%%
\be
{\bf 14}\longrightarrow (1) \quad \oplus \quad 4 \times (\ft12)\quad
\oplus\quad
3\times (0)\,,
\ee
%%%%%
and hence from (\ref{dynkin}) the Dynkin index of the $SU(2)_R$ embedding is
%%%%%
\be
I_D=1\,.
\ee
%%%%%
The two Einstein metrics we have obtained here are the ones denoted by
$G_2/SU(2)_1$ in \cite{bohker}, which were obtained in \cite{jen,Wa2}.

\subsection{$G_2/SU(2)_{\rm max}$}\label{su2maxsec}

   There is one further inequivalent 11-dimensional coset $G_2/H$ that
we may consider, for which $H$ is the maximal $SU(2)$ subgroup in $G_2$.
Under this subgroup, the adjoint decomposes as ${\bf 14}\longrightarrow
 {\bf 11} + {\bf 3}$, which, in terms of the labelling of $SU(2)$ 
representations by their spin $j$, reads
%%%%%
\be
{\bf 14}\longrightarrow (5) \quad \oplus\quad (1)\,.
\ee
%%%%%
 From (\ref{dynkin}), it follows that the Dynkin index of this embedding is
%%%%%
\be 
I_D = 28\,.
\ee
%%%%%

   We find that the (canonically-normalised) 
left-invariant 1-forms of the $SU(2)_{\rm max}$ subgroup 
are defined by the Cartan 1-form $H_{\rm max}$ and positive-root 1-form
$E_{\rm max}^+$, whose expressions in terms of (\ref{G2E}) and (\ref{G2H})
are
%%%%%
\be
H_{\rm max}= \fft1{28}\, (5 H_1 + H_2)\,,\qquad
E^+_{\rm max}= \fft1{14\sqrt2}\, (3 E_3^+ + \fft{\sqrt5}{\sqrt3}\, E_6^+)\,.
\ee
%%%%%
We accordingly find that one can make a projection into the 
11-dimensional coset metric
%%%%%
\bea
ds_{11}^2 &=& x_1\,\Big[\ft13  (\sigma_{23}- 2 \sigma_1)^2+ 
   ( \sigma_{3\hat 2} - \sigma_{2\hat 3})^2\Big] +
x_2\, \Big[\ft13(\sigma_{31}-2\sigma_2)^2+ (\sigma_{3\hat1}-\sigma_{1\hat3})^2
\Big]\,,\nn\\
&&+ x_3\, \Big[3 \sigma_{12}^2 +  (\sigma_{1\hat2}+\sigma_{2\hat1})^2\Big]
 + x_4\, \Big[3 \sigma_{31}^2 +(\sigma_{1\hat3}+\sigma_{3\hat1})^2\Big]
\,,\\
&& +x_5 \Big[ \ft13 (\ft{3\sqrt3}{\sqrt5}\, \sigma_{23} - 
                      2\sigma_3 + \sigma_{12})^2 + 
  (\ft{\sqrt3}{\sqrt5}\, (\sigma_{2\hat3}+\sigma_{3\hat2})  +
            \sigma_{1\hat2}-\sigma_{2\hat1})^2 \Big] +
  \ft43  x_6\, (2\sigma_{1\hat1}+3\sigma_{2\hat2})^2\,,\nn
\eea
%%%%%
provided that the constants are chosen (up to scale) so that 
%%%%%
\be
(x_1,x_2,x_3,x_4,x_5,x_6)=(1,1,\ft13,\ft13,\ft5{14},\ft{3}{28})\,.
\ee
%%%%%
There is no freedom, except for an overall scaling, in the choice of
the metric coefficients, the embedding is isotropy irreducible, and
thus it is Einstein \cite{wolf}.  
We find the Einstein constant and the invariant
$I_2$ are given by
%%%%%
\be
\lambda=\fft{43}{28}\,,\qquad I_2= \fft{69883}{3698}\,.
\ee
%%%%%

\subsection{$G_2/U(2)$ flag manifold}

   There are two $G_2/U(2)$ cosets that one may consider, in which the
$U(2)$ is taken to be either $SU(2)_L\times U(1)_R$, or else
$SU(2)_R\times U(1)_L$.  (The $U(1)$ factors are taken from $SU(2)_R$ or
$SU(2)_L$ respectively.) The case when $U(2)$ is $SU(2)_L\times U(1)_R$
gives rise to the flag manifold $G_2/U(2)$.

   The metric on the coset $G_2/[SU(2)_L\times U(1)_R]$ is obtained by
dividing out the $y_4$ terms and the last of the three $y_3$ terms in
(\ref{G2more}):
%%%%%
\bea
ds_{10}^2 &=& y_1\, [(\sigma_{1\hat 2} -\sigma_{2\hat 1})^2 +
 (\sigma_{1\hat 3} -\sigma_{3\hat 1})^2 +
     (\sigma_{2\hat 3} -\sigma_{3\hat 2})^2]\nn\\
&& +
 y_2\,  [(\sigma_{1\hat 2} + \sigma_{2\hat 1})^2 +
 (\sigma_{1\hat 3} + \sigma_{3\hat 1})^2 +
     (\sigma_{2\hat 3} + \sigma_{3\hat 2})^2]\nn\\
&& + y_3\, [(\sigma_{23}-\sigma_1)^2 + (\sigma_{31}-\sigma_2)^2] \nn\\
&& + y_5\, (\sigma_{1\hat1} - \sigma_{2\hat2})^2 +
     y_6\, (\sigma_{1\hat1} + \sigma_{2\hat2})^2\,.\label{G2flag}
\eea
%%%%%
This factoring can be performed provided that $y_1=y_6=3y_2=3y_5$.  The
Einstein equations then imply that $y_3=2y_1$ or $y_3=\ft23 y_1$. Thus,
up to scaling, we obtain the two inequivalent Einstein metrics
%%%%%
\bea
(y_1,y_2,y_3,y_5,y_6)=(3,1,6,1,3):&& \lambda=\ft12\,,\quad 
  I_1= \fft{27}{16}\,,\quad I_2=\fft{460}{27}\,,\nn\\
(y_1,y_2,y_3,y_5,y_6)=(3,1,2,1,3):&& \lambda= \fft{11}{18}\,,\quad
I_1= \fft{161051}{104976}\,,\quad I_2=\fft{2020}{121}\,.
\eea
%%%%%

   The coset $G_2/U(2)$ that we have constructed here, denoted by
$G_2/U(2)_3$ in \cite{bohker}, is the flag manifold of $G_2$.  (The
subscript is the Dynkin index of the $SU(2)$ factor in the denominator
subgroup.) The two
Einstein metrics were obtained first in \cite{DK2}, and were recently
discussed further in \cite{arvachry}.  
 
\subsection{$G_2/U(2)$ Grassmann manifold $G_2^+(\R^7)$}

   The other $G_2/U(2)$ coset is obtained by taking $U(2)= SU(2)_T\times
U(1)_L$, and is thus denoted by $G_2/U(2)_1$ in \cite{bohker}. As discussed
in \cite{kerr}, this $U(2)$ subgroup of $G_2$ is also contained in the
$SU(3)$ subgroup of $G_2$ (in fact it is the intersection of the $SU(3)$
and the $SU(2)\times SU(2)$ subgroups of $G_2$).  The resulting coset space is
isomorphic to the Grassmannian $G_2^+(\R^7)=SO(7)/[SO(2)\times SO(5)]$ 
of oriented 2-planes through the origin in $\R^7$ \cite{kerr}.

  The $SU(3)$ subgroup of $G_2$ is spanned by the left-invariant 1-forms 
$E^\pm_4$, $E^\pm_5$, $E^\pm_6$, $H_1$ and $H_2$ (see (\ref{G2E}) and 
(\ref{G2H})).  The $U(2)$ subgroup can be taken to be spanned by
$E^\pm_4$ and $H_1$ (spanning $SU(2)$) together with $H_2$ (spanning the
$U(1)$ factor).  Thus we may write $G_2$-invariant metrics on the Grassmannian 
$G_2/U(2)=G_2^+(\R^7)$ by dividing out the terms proportional to
$x_6$, $x_7$ and $x_8$ in (\ref{G2met12}).  This truncation is consistent 
provided that we take $x_1=x_2$ and $x_4=x_5$, and so we consider metrics
of the form
%%%%%
\bea
ds_{10}^2 &=& y_1\, \Big[ (\sigma_{2\hat 3} -\sigma_{3\hat 2})^2 +
     \ft13 (\sigma_{23} -2 \sigma_1)^2  +
 (\sigma_{3\hat 1} -\sigma_{1\hat 3})^2 +
     \ft13 (\sigma_{31} -2 \sigma_2)^2\Big] \nn\\
&&+
 y_2\, \Big[ (\sigma_{1\hat 2} -\sigma_{2\hat 1})^2 +
     \ft13 (\sigma_{12} -2 \sigma_3)^2\Big]\nn\\
&& + y_3\, \Big[3 \sigma_{23}^2 + (\sigma_{2\hat3}+\sigma_{3\hat2})^2
+   3 \sigma_{31}^2 + (\sigma_{3\hat1}+\sigma_{1\hat3})^2\Big]\,.
\eea
%%%%%
Scaling so that $y_1=1$, we find that the Einstein equations imply
%%%%%
\be
y_3= \fft{(3y_2-2)(y_2+2)}{2(5y_2^2-18y_2+8)}\,,
\ee
%%%%%
and $y_2$ must satisfy
%%%%
\be
(y_2-2)(60y_2^5 - 776 y_2^4 + 1891 y_2^3 -1570 y_2^2 + 523 y_2 -56)=0\,.
\label{six}
\ee
%%%%%
The quintic has three real roots, all of which are positive:
%%%%%
\be
y_2\approx 0.1868941\,,\qquad y_2 \approx 1.67467\,,\qquad 
y_2\approx 10.047\,.\label{3real}
\ee
%%%%%
The first two of these, and the root $y_2=2$ in (\ref{six}), all give
positive values for $y_3$ and thus yield Riemannian Einstein metrics.  These
have $\lambda$, $I_1$ and $I_2$ given by
%%%%%
\bea
(y_1,y_2,y_3)=(1,2,1):&& \lambda=\fft56\,,\qquad I_1=\fft{3125}{3888}\,,
\qquad I_2= \fft{26}{3}\,,\nn\\
(y_1,y_2,y_3)\approx(1,0.1868941,0.327159):&& \lambda\approx 1.94012\,,\nn\\
&&
I_2\approx 0.549861\,,\qquad I_2\approx 30.4872\,,\nn\\
(y_1,y_2,y_3)\approx(1,1.67467,0.684128):&& \lambda\approx 1.00414\,,\nn\\
&&
I_1\approx 0.80014\,,\qquad I_2\approx 10.8238\,.
\eea
%%%%%
(We take the volume to be $V=y_1^2 y_2 y_3^2$.)
These metrics were found in \cite{kimu,arvan}, and discussed further in
\cite{kerr}. The first metric is just the standard $SO(7)$-invariant metric
on the Grassmanian $SO(7)/[SO(2)\times SO(5)]$ \cite{kerr}.

The third root in (\ref{3real}), for which $y_3$ is negative, gives a 
pseudo-Riemannian Einstein metric of signature $(4,6)$:
%%%%%
\bea
(y_1,y_2,y_3)\approx (1,10.046978,-0.510773):&&
\lambda\approx 0.312078\,,\\
&& I_1\approx 0.007759\,,\qquad
I_2\approx -140.7999\,.\nn
\eea
%%%%%

\subsection{$G_2/SU(3) = S^6$}

   There is an $SU(3)$ maximal subgroup of $G_2$, for which the associated
left-invariant 1-forms are
%%%%%
\be
\sigma_{23}\,,\quad \sigma_{31}\,,\quad \sigma_{12}\,,\quad
(\sigma_{2\hat3}+\sigma_{3\hat2})\,,\quad
(\sigma_{3\hat1}+\sigma_{1\hat3})\,,\quad
(\sigma_{1\hat2}+\sigma_{2\hat1})\,,\quad
(\sigma_{1\hat1}-\sigma_{2\hat2})\,,\quad
(\sigma_{1\hat1}+\sigma_{2\hat2})\,.
\ee
%%%%%
We find that there is a unique (up to overall scale) Einstein metric on
$G_2/SU(3)$, given by
%%%%%
\bea
ds_6^2 &=&  (\sigma_{2\hat 3} -\sigma_{3\hat 2})^2 +
(\sigma_{3\hat 1} -\sigma_{1\hat 3})^2 +
(\sigma_{1\hat 2} -\sigma_{2\hat 1})^2\nn\\
&& +\ft13 (\sigma_{23} -2 \sigma_1)^2 +
  \ft13 (\sigma_{31} -2 \sigma_2)^2 +
   \ft13 (\sigma_{12} -2 \sigma_3)^2\,.
\eea
%%%%%
This is $S^6$, with its standard Einstein metric. (With the 
scaling we have chosen, it has $\lambda= 5/3$.)

\section{Conclusion}

   In this paper, we have found four inequivalent positive-definite
Einstein metrics on
the group manifold $SO(5)$, and six inequivalent positive-definite
Einstein metrics on
$G_2$.  Two of the metrics on $SO(5)$, and four of the metrics on $G_2$,
appear to be new.  One motivation for studying this question was the possible 
utility of such metrics for the construction of background solutions in
supergravity, string theory and M-theory.  

    Mindful of the possible applications to the Chronology Protection 
Conjecture, and related issues concerning closed timelike curves (CTCs),
we also searched for Einstein metrics of indefinite signature.  We found
one on $SU(3)$ with signature $(2,6)$, one on $G_2/SU(2)_{\rm diag}$
with signature $(6,5)$, and one on the Grassmannian $G_2^+(\R^7)=G_2/U(2)$
with signature $(4,6)$.  The absence of Lorentzian examples is
striking, since there is no topological obstruction to a group manifold, 
compact or otherwise, admitting a Lorentzian metric (although it will, if
compact, have CTCs).  Indeed, the necessary and sufficient condition
that a manifold admit a time orientable Lorentzian metric is that it
admit an everywhere non-vanishing vector field, or, equivalently, that the
Euler number vanish.  This condition holds trivially for group manifolds,
and for all odd-dimensional compact manifolds.  Of course, if we relaxed
the Einstein condition it would be trivial to write down Lorentzian 
metrics on group manifolds, simply by taking the matrix $x_{ab}$ in 
(\ref{sqmet}) to have one negative eigenvalue. 

  It is possible that the absence of Lorentzian metrics may be ascribed to
the restricted nature of our ans\"atze.  It may also be, by analogy with the
G\"odel solution, that to obtain Lorentzian metrics satisfying the
Einstein equations, one needs to add material sources such as a perfect
fluid.  This is an interesting topic for future investigation.

\end{document}